\documentclass[reprint,english,superscriptaddress,preprintnumbers,amsmath,amssymb,aps,prx]{revtex4-1}

\usepackage{diagbox}

\usepackage{mathtools}
\usepackage{amsfonts}
\usepackage{mathrsfs}
\usepackage{bbm}
\usepackage{slashed}

\usepackage{graphicx}
\usepackage{color}
\usepackage{array}

\usepackage{blindtext}
\usepackage{placeins}
\usepackage{booktabs}
\usepackage{makecell}
\usepackage{epstopdf}
\usepackage[caption=false]{subfig}
\usepackage{xspace}
\usepackage{siunitx}
\usepackage{hyperref}
\usepackage[nameinlink]{cleveref}
\usepackage{appendix}

\usepackage{xifthen}
\usepackage{xcolor}
\hypersetup{
	colorlinks,
	linkcolor={red!75!black},
	citecolor={blue!75!black},
	urlcolor={blue!75!black}
}

\setkeys{Gin}{width=0.48\textwidth}

\graphicspath{
	{./figures/}
	{../figures/}
}

\def\s0#1#2{\mbox{\small{$ \frac{#1}{#2} $}}}
\def\0#1#2{\frac{#1}{#2}}

\newcommand{\sumint}{\int\hspace{-4.8mm}\sum}
\newcommand{\imag}{\text{i}}

\usepackage{multirow}
\usepackage{colortbl}
\definecolor{kugray5}{RGB}{224,224,224}
\newcommand{\PreserveBackslash}[1]{\let\temp=\\#1\let\\=\temp}
\newcolumntype{C}[1]{>{\PreserveBackslash\centering}p{#1}}
\newcolumntype{R}[1]{>{\PreserveBackslash\raggedleft}p{#1}}
\newcolumntype{L}[1]{>{\PreserveBackslash\raggedright}p{#1}}

\begin{document}

\title{QCD equation of state and thermodynamic observables \\ from computationally minimal Dyson-Schwinger Equations}

\author{Yi Lu}
\email{qwertylou@pku.edu.cn}
\affiliation{Department of Physics and State Key Laboratory of Nuclear Physics and Technology, Peking University, Beijing 100871, China}

\author{Fei Gao}
\email[Corresponding author: ]{fei.gao@bit.edu.cn}
\affiliation{School of Physics, Beijing Institute of Technology, 100081 Beijing, China}

\author{Yu-xin Liu}
\email{yxliu@pku.edu.cn}
\affiliation{Department of Physics and State Key Laboratory of Nuclear Physics and Technology, Peking University, Beijing 100871, China}
\affiliation{Center for High Energy Physics, Peking University, 100871 Beijing, China}
\affiliation{Collaborative Innovation Center of Quantum Matter, Beijing 100871, China}

\author{Jan M. Pawlowski}
\email{J.Pawlowski@thphys.uni-heidelberg.de}
\affiliation{Institut f{\"u}r Theoretische Physik,
	Universit{\"a}t Heidelberg, Philosophenweg 16,
	69120 Heidelberg, Germany
}
\affiliation{ExtreMe Matter Institute EMMI,
	GSI, Planckstr. 1,
	64291 Darmstadt, Germany
}

\begin{abstract}
	
We study the QCD equation of state and other thermodynamic observables including the isentropic trajectories and the speed of sound. These observables are of eminent importance for the understanding of experimental results in heavy ion collisions and also provide a QCD input for studies of the timeline of heavy-ion-collisions with hydrodynamical simulations. They can be derived from the quark propagator whose gap equation is solved within a minimal approximation to the Dyson-Schwinger equations of QCD at finite temperature and density. This minimal approximation aims at a combination of computational efficiency and simplification of the truncation scheme while maintaining quantitative precision. This  minimal DSE scheme is confronted and benchmarked with results for correlation functions and observables from first principles QCD lattice at vanishing density and quantitative functional approaches at finite density.

\end{abstract}

%\pacs{25.75.Nq, 11.10.Wx, 12.38.Lg, 21.65.Qr }

\maketitle

%%%%%%%%%%%%%%%%%%%%%%%%%%%
\section{introduction}
\label{sec:Intro}

%Dropped References: \cite{ Dupuis:2020fhh, Fu:2022gou}.

The thermodynamic properties of strong interaction matter are of both experimental and theoretical interest.
The phase structure of strongly interacting matter is explored in  currently running  and planned on-going heavy-ion-collision facilities such as the BNL Relativistic Heavy Ion Collider (RHIC), GSI Facility for Antiproton and Ion Research (FAIR), JINR Nuclotron-based Ion Collider facility (NICA) and High Intensity heavy ion Accelerator Facility (HIAF).
Its thermodynamic properties in the phase structure are governed by the QCD equation of state (EoS), i.e. thermodynamic functions such as pressure, entropy density, energy density, etc., at finite temperature $T$ and quark chemical potential $\mu_q$~\cite{Klevansky:1992qe, Buballa:2003qv, Fukushima:2013rx, Fukushima:2017csk}. Specifically, for hydrodynamic simulations of heavy-ion collision, the QCD EoS is a crucial input as are further transport coefficients, see e.g.~\cite{Freedman:2013awl,Rischke:2003mt}. Moreover, at large densities and small temperatures, the QCD EoS is required for explaining the physics of compact stars such as neutron stars, e.g.~\cite{Oertel:2016bki}.

Accordingly, obtaining the EoS and other thermodynamic observables from first principles QCD is of utmost importance for the physics phenomena discussed above. At finite chemical potential and in particular for $\mu_B/T \gtrsim 3$ these results can only be obtained with functional QCD approaches such as Dyson-Schwinger equations (DSE) and the functional renormalisation group (fRG) approach, as lattice simulations at finite chemical potential to date suffers from the sign problem. Investigations of the phase structure of QCD with functional QCD approaches have made significant progress over the past decade, see in particular~\cite{Qin:2010nq, Fischer:2012vc, Fischer:2014ata, Fu:2019hdw, Gao:2020fbl}, and the reviews \cite{Roberts:2000aa, Fischer:2018sdj} (DSE) and \cite{Dupuis:2020fhh, Fu:2022gou} (fRG). In turn, at vanishing $\mu_B$, first principles QCD computations on the lattice provide benchmark results for the chiral phase transition temperature, thermodynamic observables and fluctuations of conserved charges. QCD, see e.g.~\cite{Borsanyi:2020fev, HotQCD:2018pds, Bonati:2018nut}, which also can be used for extrapolations to finite chemical potential~\cite{Borsanyi:2012cr,Bazavov:2017dus,Borsanyi:2021sxv}.

By now the results for the chiral phase structure from functional approaches are converging quantitatively at finite density with the increasing order of the truncations used. Moreover, these up-to-date results meet the lattice benchmark results at vanishing (and low) chemical potential, see \cite{Fu:2019hdw, Gao:2020qsj, Gao:2020fbl, Gunkel:2021oya}. The convergent results include an estimate for an onset regime of new physics, potentially a CEP, at about $(T,\mu_B) \sim (110\,,\,600)\,$MeV. This location lies beyond the quantitative convergence regime $\mu_B/T\lesssim 4$ of the current best approximations, and hence it is only an estimate and not a fully qunatitative prediction. Still, this is exciting news and furthermore there is an ongoing quest for even more elaborate truncations that aim for full apparent convergence. However, the present approximations already allow for quantitative computation in the regime $\mu_B/T\lesssim 4$ and for estimates in the regime $\mu_B/T\gtrsim 4$.

This opens the path towards a comprehensive analysis of the equation of state, further thermodynamic observables, fluctuations of conserved charges as well as timelike observables such as transport coefficients within functional approaches. In the present work we contribute twofold to this endeavour: \\[-2ex]

\textit{(i)} We want to make quantitative functional QCD computations accessible to a wider audience beyond the technical experts. To that end we set up a minimal computational scheme for DSE computation: such a scheme aims at the technically most simple approximation  at finite temperature and density that still reproduces the phase structure results with the state-of-the-art approximation scheme in \cite{Fu:2019hdw, Gao:2020qsj, Gao:2020fbl, Gunkel:2021oya} and hence allows a relatively simple access to many observables beyond the phase structure itself.   \\[-2ex]

\textit{(ii)} We compute the equation of state and other thermodynamic observables  in a wide range of $T$ and $\mu_B$ within this scheme. This allows us to study further thermodynamic observables such as the isentropic trajectories and the speed of sound, highly relevant for hydrodynamic simulations at finite density. \\[-2ex]

This work is organised as follows: In \Cref{sec:miniDSE}, we present the framework of the minimal scheme and its agreement with the other studies in vacuum. In \Cref{sec:PhaseStructure},  we apply the framework in the plane of temperature and chemical potential and obtain the chiral phase transition. Then in \Cref{sec:EoS}, we present the results of EoS in the ($\mu$, $T$) plane and also the isentropic trajectories.  In \Cref{sec:Summary}, we summarise the main results and make further discussions and outlook.

%%%%%%%%%%%%%%%%%%%%%%%%%%%%%%%%%%%%
\section{The minimal DSE scheme}
\label{sec:miniDSE}

\begin{figure}[t]
	\centering
	\includegraphics[width=0.95\columnwidth]{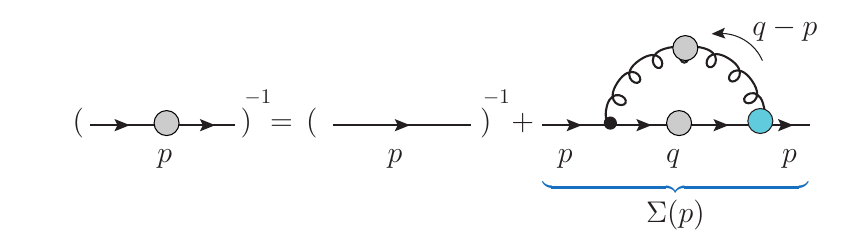}
	\caption{Quark DSE for the quark self energy $\Sigma(p)$.}
	\label{fig:QuarkDSE}
\end{figure}
In this section we develop the minimal truncation scheme for the DSE approach at finite temperature and density, that is
minimal for quantitative and semi-quantitative results with minimal computational effort (miniDSE). Key to this approach is the quantitative solution of the quark gap equation or quark DSE for the full quark propagator $S(p)$,
\begin{subequations}
	\label{eq:QuarkDSE}
\begin{align}
	\left[S(p)\right]^{-1} = \left[S_0(p)\right]^{-1} + \Sigma(p)\,,
\label{eq:QuarkDSE1}
\end{align}
where $S_0$ is the classical propagator,
\begin{align}
	S_0(p) =\frac{1}{\imag \slashed{p} +m}\,,
	\label{eq:Sf0}
\end{align}
where $m$ is the matrix of current quark masses with entries $m_f$ for all flavours $f=1,...,N_f$. $\Sigma$ is the renormalised self energy, that satisfies the DSE
\begin{align}
	\Sigma(p) \simeq \frac{4}{3} \, g_s \, \sumint\limits_q \, \gamma_\mu S(q)\Gamma_{\nu}(q,p)D_{\mu\nu}(k)\,,
\label{eq:QuarkDSE2}
\end{align}
\end{subequations}
where we have dropped the renormalisation details. The diagrammatic depiction of \labelcref{eq:QuarkDSE} is provided in \Cref{fig:QuarkDSE} and the momentum arguments in the quark-gluon vertex are the incoming quark and antiquark momenta. \Cref{eq:QuarkDSE2} is computed within the MOM${}^2$ scheme developed in \cite{Gao:2020fbl, Gao:2020qsj, Gao:2021wun} also used implicitly in fRG computations. We refer to \cite{Gao:2021wun} for a detailed analysis of this RG-scheme in the vacuum.

In \labelcref{eq:QuarkDSE2}, the gluon momenta $k$ of the full gluon propagator $D_{\mu\nu}$ is given by $k = q - p$, and the classical quark propagator $S_{0}(p)$ in \labelcref{eq:Sf0} is flavour-diagonal. The full quark-gluon vertex $\Gamma_{\nu}$ is also taken flavour-diagonal, and hence \labelcref{eq:QuarkDSE2} constitutes equations for the self-energies $\Sigma^{f}$ for a given flavour $f$, that only depends on the classical  and full quark propagators $S^{f}_0\,,\,S^{f}$ or rather $\Sigma^f$ of the same flavour $f$ and $\Gamma_{\nu}^{f}$. Hence, the quark gap equation is flavour-diagonal, however, the gluon propagator depends on all flavours.

In the current work we restrict ourselves to 2+1 flavour QCD with $f=u,d,s$. In the vacuum, the full quark propagator is parameterised with a flavour diagonal Dirac dressing $A$ and a scalar dressing $B$, to wit,
\begin{align}
  S^{-1}(p) = i \slashed{p} \, A(p) + B(p)\,.
  \label{eq:Sp}
\end{align}
The vacuum gluon propagator is transverse in the Landau gauge used in the current work and has the transverse dressing $Z$,
\begin{align}
  D_{\mu\nu}(k) =G_A(k)   \Pi^\bot_{\mu\nu}(k) \,, \quad
  \Pi^\bot_{\mu\nu}(k) = \delta_{\mu\nu} - \frac{k_{\mu} k_{\nu}}{k^2}\,,
 \label{eq:gluon}
 \end{align}
with the transverse projection operator $\Pi^\bot$ and the scalar propagator part
\begin{align}
G_A(k)= 	 \frac{Z(k)}{k^2} \,.
\label{eq:GA}
\end{align}
The vacuum gluon propagator  \labelcref{eq:gluon} is not computed in the present work, as by now there are very accurate and consistent results for the 2+1 flavour gluon propagator from lattice QCD simulations and functional computations~\cite{RBC:2014ntl, Zafeiropoulos:2019flq, Fu:2019hdw, Gao:2020qsj, Gao:2020fbl}. Therefore we use the parametrised formula of the $2+1$ flavour gluon put forward in~\cite{Gao:2021wun}.

The last ingredient is the quark-gluon vertex. In the vacuum it can be built from eight transverse tensor structures $\{{\cal T}^{(i)}\}$ with $i=1,..,8$ and 4 longitudinal ones, see e.g.~\cite{Gao:2021wun}. For the development of our  minimal truncation scheme at finite density and temperature we can build on many functional results obtained for the quark-gluon vertex in the vacuum, see e.g.~\cite{Chang:2010hb, Williams:2014iea, Mitter:2014wpa, Williams:2015cvx, Cyrol:2017ewj, Tang:2019zbk, Gao:2021wun, Chang:2021vvx}.

%%%%%%%%%%%%%%%%%%%%%%%%%%%%%%%%
\subsection{Minimal Scheme for functional approaches}
\label{sec:miniFun}

Here we put forward a minimal scheme in functional approaches (miniDSE or miniFRG) that allows for quantitative results with a small systematic error. It builds on previous developments in \cite{Fu:2019hdw, Gao:2020qsj} and is built on two pillars:\\[-1ex]

\textit{(i) Minimal fluctuations:} it is an advantageous property of functional approaches such as the DSE and the fRG that QCD correlation functions such as continuum extrapolated lattice results or quantitative functional QCD results can be implemented straightforwardly. Moreover, functional loop equations of QCD for given external parameters such as temperature $T$, baryon chemical potential $\mu_B$, and number of quark flavours $N_f$ can be expanded about QCD for different external parameters, for more details see \cite{Gao:2020qsj, Gao:2020fbl}. This minimises the amount of quantum, thermal and density fluctuations carried by the functional equations themselves. The benefits of such a procedure are twofold: Firstly, assuming a negligible or small systematic error of the input  it minimises the systematic error as the latter only concerns the fluctuations carried by the functional equations. This allows us to reduce the intricacy of the approximation within the DSE or fRG considerably without a significant loss of the quantitative nature of the result. Secondly, it minimises the need of renormalising the functional equations. In the DSE the latter is highly non-trivial within non-perturbative approximations, while the gain is the fRG is the qualitative reduction of UV-relevant running with positive powers of the cutoff scale.

We exemplify the procedure within the quark gap equation,
\begin{align}
	\Sigma_{\boldsymbol{v}} = 	\Sigma_{\boldsymbol{v}'} +\Delta \Sigma\,,
\label{eq:ExpandS}
\end{align}
with the difference of the self energies  	
\begin{align}
\Delta \Sigma_{\boldsymbol{v},\boldsymbol{v}'}  = 	\Sigma_{\boldsymbol{v}} -	\Sigma_{\boldsymbol{v}'} \,,
\label{eq:DeltaS}
\end{align}	
and $\boldsymbol{v}$ collects the external parameters, e.g.~$\boldsymbol{v}=(T,\mu_B,N_f,m_f)$.
Equating \cref{eq:DeltaS} with the difference of the DSEs constitutes a closed gap equation for $\Delta \Sigma$ with the input $\Sigma_{\boldsymbol{v}'}$ and
the quark-gluon vertices $\Gamma_{\mu\boldsymbol{v}}$ and $\Gamma_{\mu;_{\boldsymbol{v}}'}$. Evidently, the closer ${\boldsymbol{v}}$ is to ${\boldsymbol{v}}'$, the less non-trivial physics is implemented by the loop itself. Moreover, for $(N_f,m_f)=(N_f',m_f')$ the difference DSE is finite and does not require renormalisation.\\[-1ex]

\textit{(ii) Minimal correlation functions:} Complete $n$-point correlation functions $\Gamma^{(n)}(p_1,...,p_n)$ carry a rapidly increasing number of tensor structures. Their respective scalar dressing functions, which depend on the momenta $p_1,...,p_{n-1}$, the remaining momentum $p_n$ is fixed by momentum conservation. However, only few of these dressings have a sizable impact within the system of functional equations and higher order vertices are typically suppressed due to space-time and momentum locality of the vertices in gauge-fixed QCD, for more details see \cite{Dupuis:2020fhh}. In our example of the difference gap equation for \labelcref{eq:ExpandS} a respective evaluation concerns only the quark-gluon vertex $\Gamma_\mu$ or rather its difference
\begin{align}
	\Gamma_{\mu;\boldsymbol{v}} = 	\Gamma_{\mu;\boldsymbol{v}'} +\Delta \Gamma_{\mu;{\boldsymbol{v},\boldsymbol{v}'} }\,, \qquad \Delta \Gamma_{\mu;\boldsymbol{v},\boldsymbol{v}'}  = 	\Gamma_{\mu;\boldsymbol{v}} -	\Gamma_{\mu;\boldsymbol{v}'} \,,
	\label{eq:DeltaGammamu}
\end{align}
which is the main external ingredient in the gap equation. In a first application of the minimal scheme or miniDSE we will construct a reduced minimal truncation of the quark-gluon vertex with only two tensor structures in \Cref{sec:miniQuarkGluon}. In general such a construction uses the  space-time and momentum locality of the vertices as well as benchmark results within full functional computations and lattice simulations. \\[-1ex]

In summary, the above minimal scheme allows us to obtain quantitatively reliable results for observables with a significant reduction of the numerical costs and a sizable improvement of the stability of the convergence of the numerics. In combinations this can lead to a reduction of the computation time by orders of magnitude. Moreover, some of these reduced truncations in the miniDSE are easily accessible technically also for non-experts.

%%%%%%%%%%%%%%%%%%%%%%%%%
\subsection{Quark-gluon vertex in the miniDSE scheme}
\label{sec:miniQuarkGluon}

The quark-gluon vertex in the vacuum has a complete basis of twelve tensor structures, and its transverse part can be expanded in eight transverse projections of these tensors.  At finite temperature and density these transverse projections all come with a thermal split.

The following suggestion for a simplified four-quark vertex in the miniDSE scheme builds on results of the in-detail analysis of the  importance ordering of the vertices in \cite{Gao:2021wun} in the vacuum. Moreover we also work in the information from DSE results at finite temperature and density obtained in the precursor of the present minimal scheme in \cite{Gao:2020qsj}, and its comparison with the full computation in \cite{Gao:2020fbl}. This combined analysis showed that five of the eight tensor structures are completely irrelevant and we only have to consider the transverse projections of the remaining three,
\begin{align}\nonumber
{\cal T}_\mu^{(1)}(q,p) = &\, -\imag \gamma_\mu \,,\\[1ex]\nonumber 	
		{\mathcal{T}_\mu^{(4)}}(q,p)
	= &\, -\imag \sigma_{\mu\nu} k^{\nu}\, \,,\qquad \sigma_{\mu\nu} = \frac{\imag}{2} \, [\gamma_{\mu},\gamma_{\nu}]\,, \\[1ex]
	{\mathcal{T}_\mu^{(7)}}(q,p) =& \, \frac{i}{3} \Big\{ \sigma_{\alpha\beta}\gamma_{\mu} + \sigma_{\beta\mu}\gamma_{\alpha} + \sigma_{\mu\alpha}\gamma_{\beta} \Big\} l^{\alpha} k^{\beta} \,,
\label{eq:Tensors147}
\end{align}
each coming with a momentum dependent dressing function  $\lambda^{(i)}(q,-p)$ with the incoming quark and antiquark momenta $q$ and $-p$  respectively, and the gluon momentum $k_\mu$ and the weighted sum of the quark and antiquark momenta $l_\mu$,
\begin{align}
	k=q-p\,, \qquad \qquad l=\frac12 \bigl( p+q\bigr)\,.
\label{eq:gluonMom}
\end{align}
Then, the miniDSE quark-gluon vertex takes the form
\begin{align}
	\Gamma_\mu(q,p) = \sum_{1,4,7} {\cal T}_{\mu}^{(i)}(q,p) \lambda^{(i)}(q,p)\,.
\label{eq:miniDSEquarkgluon}
\end{align}
The terms in \labelcref{eq:miniDSEquarkgluon} have the following relevance ordering~\cite{Gao:2021wun}: the by far dominant component of the vertex is that with the classical (chiral) tensor structure, ${\cal T}_1\,\lambda_1$, and the dressing is constrained by the Slavnov-Taylor identities (STIs). This is followed by the chiral symmetry breaking part ${\cal T}_4\,\lambda_4$. The smallest contribution originates in the second chirally symmetric part ${\cal T}_7\,\lambda_7$. The Dirac structures of quark-gluon vertex  are adopted from  \cite{ Gao:2021wun}, except $\mathcal{T}_7$ which has less overlap with the other components and avoids kinematic singularities due to its symmetric form, see~\cite{Eichmann:2016yit}.

Then, the fully quantitative miniDSE scheme would utilize the splits   \labelcref{eq:ExpandS} and \labelcref{eq:DeltaGammamu} with $\boldsymbol{v}'=(N_f',m_f',T',\mu_B') = (N_f,m_f,0,0)$ or even with $T'=T$ as well as the quantitative data from \cite{Gao:2021wun} or finite temperature results. Moreover, at finite temperature and density the dressings $\lambda_{1,4,7}$ with and without thermal split would be approximated by combinations of the dressings of the quark propagator as done in \cite{Gao:2020qsj}. The latter step further reduces the numerical costs significantly. The quantitative nature of this approximation has already been confirmed in \cite{Gao:2020qsj, Gao:2020fbl}. This concludes our discussion of the quantitative miniDSE scheme for applications to the phase diagram of QCD.

In the present work we will further simplify the scheme by approximating the vertex dressings also at $T=0$ with combinations of the propagator dressings. Moreover, we shall drop the least important part ${\cal T}_7 \lambda_7$, even though it accounts for an about  20\% decrease of the mass function. We accommodate for this decrease of the mass function by decreasing the coupling constant with roughly 3\% compared with the full QCD coupling in \cite{Gao:2021wun}. We emphasize that this is based on a self-consistency check of the quantitative nature of the procedure, checked with the full results also at finite temperature and chemical potentials relevant for the chiral phase structure and thermodynamic observables studied here.

In summary this leads us to a computationally minimal scheme only in terms of the quark dressings with the quark gluon vertex
\begin{align}
	\Gamma_{\mu}(q,p) ={\cal T}^{(1)}_\mu(q,p) \,\lambda^{(1)}(q,p)  + {\mathcal{T}^{(4)}_{\mu}}(q,p)\,\lambda^{(4)}(q,p)\,,
	\label{eq:Minimalqgvertex}
\end{align}
where the dressing of the classical tensor structure is constrained by the STIs for the quark-gluon vertex. We shall use
\begin{align}
	\lambda^{(1)}(q,p) = g_s  F(k^2) \Sigma_A(q,p)\,,
	\label{eq:lambda1}
\end{align}
with the ghost dressing function $F(k^2)=k^2 \, G_c(k)$, where $G_c(k) \delta^{ab}$ is the ghost propagator. The other factor $\Sigma$ is the sum of the quark dressings $A$ defined in \labelcref{eq:Sp},
\begin{align}
\Sigma_A(q,p) = \frac{A(p)+A(q)}{2}\,.
\label{eq:Sigma}
\end{align}
Several studies suggest that $\lambda^{(4)}$ is proportional to differences of the scalar quark dressing function~\cite{Chang:2011ei, Qin:2013mta, Gao:2016qkh},
\begin{align}
  \Delta_B(q,p) = \frac{B(p)-B(q)}{p^2-q^2}\,.
\label{eq:DeltaB}
\end{align}
The scalar dressing of the quark propagator carries the RG-scaling of the quark and anti-quark leg of the quark-gluon vertex. The RG-scaling of any vertex dressing $\lambda^i$ also has to accommodate the RG-scaling of the gluon leg $\propto 1/Z^{1/2}(k)$ with the gluon dressing defined in \labelcref{eq:gluon}.  It has been shown in~\cite{Gao:2020qsj, Gao:2021wun} by comparison to the full vertex computed in \cite{Gao:2021wun} (DSE) and fRG \cite{Cyrol:2017ewj} (fRG) in the MOM${}^2$ scheme, that this factor indeed not only carries the appropriate RG-scaling but also the correct momentum dependence of $\lambda_4$ in the vacuum. Hence, in the vacuum we choose
\begin{align}
\lambda^{(4)}(q,p) = \frac{g_s}{ Z^{1/2}(k)} \, \Delta_B(q,p)\,,
\label{eq:lambda_4}
\end{align}
with $Z(k)$ the gluon dressing function introduced in \labelcref{eq:gluon}, see \cite{Gao:2020qsj}. \Cref{eq:lambda_4} introduces a kinematic singularity into the vertex that it absent in the direct computation. Note however, that in our computations, the vertex is always attached to a gluon propagator with momentum $k$ and the factor $1/Z^{1/2}(k)$ is cancelled. Moreover, the loop integration introduces a further $k^2$ at finite temperature and $k^3$ in the vacuum, which leads to a very efficient suppression of this regime. This is checked with a comparison to the results from computations with full vertices which allows for a systematic error estimate.

As a part of this evaluation we first argue that the kinematic singularity can be avoided by the following upgrade of the present procedure: Instead of using \labelcref{eq:gluon} and its finite temperature and chemical potential analogues for the definition of the gluon wave function, one can use a parameterisation for the scalar propagator part $G_A(k)$ in \labelcref{eq:gluon}, that takes into account the mass gap of QCD explicitly. In the vacuum this reads
\begin{align}
		G_A(k)=\frac{1}{ Z_{A,\textrm{scr}}(k) }\frac{1}{k^2 +m_\textrm{scr}^2}\,,
		\label{eq:GAZA}
\end{align}
where $m_\textrm{scr}^2$ is the spatial screening mass. This mass is defined via the exponential decay of the large distance limit of the spatial Fourier transform of the gluon propagator,
\begin{align}
	\tilde G_A(k_0,r) = \int \frac{ {d}^3 k }{(2 \pi)^3}\,G_A(k_0,\boldsymbol{k} )\, e^{\imag  \boldsymbol{k} \boldsymbol{x}}\,,
\end{align}
with the spatial momentum $\boldsymbol{k}$ and the spatial position or distance $\boldsymbol{x}$ and $r=\|\boldsymbol{x}\|$. The large distance limit $r\to \infty$ can be parametrised with
\begin{align}
	\lim_{r\to \infty} \tilde G_A(k_0=0,r) \to R(r) \,e^{- m_\textrm{scr} r}\,.
\label{eq:tildeGA}
\end{align}
where $R(r)$ is a polynomial or at most a rational function of $r$. The spatial screening mass $m_\textrm{scr}^2$ is the inverse screening length and is defined as the strength of the exponential decay. A similar definition holds true for the temporal screening mass, that is obtained from the asymptotic time-dependence of the Schwinger function.

In the vacuum these two masses agree due to Lorentz invariance and we get from the functional and lattice 2+1 gluon data in  \cite{Fu:2019hdw, Gao:2020qsj, Gao:2020fbl, Boucaud:2018xup, Zafeiropoulos:2019flq},
\begin{align}
 	m_\textrm{scr} \approx 850\,\textrm{MeV}\,.
 	\label{eq:mscr}
\end{align}
The overall error of \labelcref{eq:mscr} and the respective ones for $N_f=2$ flavour QCD and Yang-Mills theory is about 20\,MeV which can be reduced significantly if producing dedicated data for the task of determining the screening mass. \Cref{eq:mscr} can be considered as a physics definition of the gluon mass gap, and can be compared with $m_\textrm{scr} \approx 830\,\textrm{MeV}$ for the two-flavour data from \cite{Cyrol:2017ewj} that underlie the 2+1 flavour computations in \cite{Fu:2019hdw, Gao:2020qsj, Gao:2020fbl} and $m_\textrm{scr} \approx 760\,\textrm{MeV}$ in Yang-Mills theory from the gluon data in  \cite{Cyrol:2016tym}, compatible with the $T\to 0$ extrapolation of the the finite temperature screening mass computed in  \cite{Cyrol:2017qkl}. The physical nature of this definition is corroborated by the quantitative agreement of the screening mass with the Debye screening mass in thermal perturbation theory for temperatures $T \gtrsim 2 T_c$, where $T_c$ is the critical temperature of the confinement-deconfinement phase transition.

The spatial and temporal screening masses differ at finite temperature and chemical potential, and a more quantitative vertex construction at finite temperature and chemical potential takes into account both screening masses. For a respective discussion and computation in finite temperature Yang-Mills theory see \cite{Cyrol:2017qkl}, and the notation in \labelcref{eq:GAZA} is close to that used there and in further fRG works such as \cite{Fu:2019hdw} and the DSE works \cite{Gao:2020qsj, Gao:2020fbl, Gao:2021wun, Gao:2021vsf}.

We emphasise that the spatial and temporal screening masses reflect the physical gluon mass gap in QCD even in the present gauge-fixed settings and constitute a relevant physics input in phenomenological considerations in the phase structure of QCD. This is already evident for its importance for the confinement-deconfinement phase transition in Yang-Mills theory, see  \cite{Cyrol:2017qkl}. Importantly, with the substitution
\begin{align}
	Z^{1/2}(k) \to \frac{1}{Z_{A,\textrm{scr}}^{1/2}(k)}\,,
	\label{eq:Z-Zscr}
\end{align}
in \labelcref{eq:lambda_4} as well as other dressings, kinematic singularities are avoided and the respective dressings reflect the decoupling of the dynamics below the (gluon) mass gap of QCD. This as well as their phenomenological importance will be considered elsewhere.

For the present purposes we find that the simplified vertex construction \labelcref{eq:lambda_4} serves well and the kinematic singularity has no impact on the physics considered here. We proceed with the systematic error estimate with a comparison to results with the full vertex. First we note, that the negligible impact of this kinematic singularity has been discussed in detail in \cite{Gao:2020qsj}, based on the explicit vacuum results in \cite{Mitter:2014wpa, Williams:2014iea, Williams:2015cvx, Cyrol:2017ewj, Gao:2021wun}. Importantly, this analysis  has also been extended to finite $T$ and $\mu_B$ in \cite{Gao:2020fbl}. Below we briefly discuss these different checks:

In \cite{Gao:2020qsj}, it has been shown, that \labelcref{eq:lambda_4} describes the full vertex in the vacuum very well down to momenta $k\approx 1$\,GeV, using also vertex data from \cite{Cyrol:2017ewj}. This has later been corroborated with vertex data from the quantitative DSE vacuum computation in \cite{Gao:2021wun}. In turn, for $k\lesssim 500$\,MeV, the vertex \labelcref{eq:lambda_4} shows a kinematic singularity which is not present in the full vertex that monotonously rises and approaches a constant for $k=0$. The  kinematic singularity in \labelcref{eq:lambda_4} is in a regime which is suppressed by the mass gap of QCD, and hence it has no impact. This has been checked and confirmed in several ways: Its reliability for computations in the phase structure has been benchmarked with the good agreement of the results with that from \cite{Fu:2019hdw} up to baryon chemical potentials $\mu_B \lesssim 600$\,MeV, and this has been corroborated by the phase structure results with the full quark-gluon vertex in the DSE computation in \cite{Gao:2020fbl}. In the present work we check the irrelevance of the kinematic singularity by freezing $Z(k)$ in \labelcref{eq:lambda_4} for small momenta with a freezing scale  in the regime
\begin{align}
k_\textrm{freeze} \approx 0.4 - 1.7\,\textrm{GeV}\,,
\end{align}
which is roughly $1/2 m_\textrm{scr} \lesssim k_\textrm{freeze}\lesssim 2 m_\textrm{scr} $ with the 2+1 flavour screening mass in \labelcref{eq:mscr}. This emulates the effect of \labelcref{eq:Z-Zscr} as $Z_{A,\textrm{scr}}$ indeed freezes for small momenta. Moreover, it covers efficiently the difference to the full vertex: while the kinematic singularity leads to an enhancement of the vertex, the freezing leads to a lowering of the vertex in comparison to the full vertex. The results do not change by more than 3\,$\%$,  which is well within the systematic error estimate of our computation and hence supports our procedure.

This concludes the discussion of the simplified version of the miniDSE scheme in the quark sector used in the present paper. The price to pay for this last simplification steps \labelcref{eq:Minimalqgvertex,eq:lambda_4} already at $T,\mu_B=0$ is a loss of quantitative reliability for baryon chemical potentials with $\mu_B/T\gtrsim 3$. This loss of quantitative reliability manifests itself e.g.~in an increasing difference of the chiral crossover line from that in full QCD in \Cref{fig:PD} for these chemical potentials including a 10\% reduction of the temperature and the chemical potential values of the location of the critical end point from the estimated regime in full quantitative functional QCD.

\begin{figure}[t]
  \centering
  \includegraphics[width=0.45\textwidth]{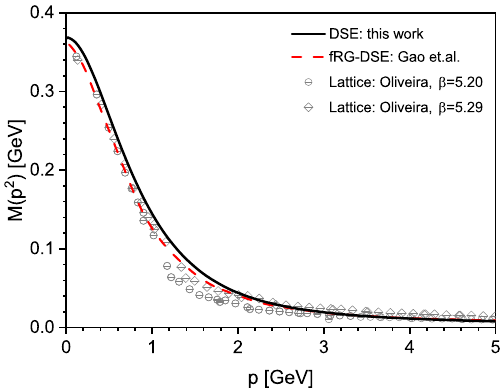}
  \caption{Light ($u,d$) quark mass functions $M(p^2) = B(p^2)/A(p^2)$ calculated from the truncation scheme \labelcref{eq:Minimalqgvertex}. The results from lattice QCD simulation~\cite{Oliveira:2016muq}
  and previous  DSE  computation in the fully coupled scheme~\cite{Gao:2021wun}  are also shown for comparison.}\label{fig:unquench}
\end{figure}
The vertex dressings in \labelcref{eq:Minimalqgvertex,eq:lambda_4} are also based on dressings from the ghost-gluon sector. The ghost propagator is almost independent of temperature and density and we use the vacuum fRG data in two-flavour QCD~\cite{Cyrol:2017ewj}. In turn, the
gluon dressings are computed from a difference DSE analogously to that of the quark discussed around \labelcref{eq:DeltaS}. The respective difference DSE have been discussed in detail in \cite{Gao:2020qsj, Gao:2020fbl}. This procedure accommodates further intricacies that arise from the need of a numerically optimal treatment of differences of frequency integrals and Matsubara sums, and hence we defer its description to the next section, \Cref{sec:EoS}, where the setup at finite $T,\mu_B$ is described, see %
\labelcref{eq:glqusplit,eq:DeltaPiAqu,eq:PiAqu,eq:glDSE2}.
%\labelcref{eq:glDSE,eq:glDSE2}.

With this input and simplification of the miniDSE scheme, the quark propagators are computed in the isospin symmetry approximation with $m_u=m_d=m_l$ with the coupling parameters $\alpha_s=g_s^2/(4 \pi),m_l,m_s$ being fixed at an RG-scale $\mu= 15$\,GeV. This is significantly lower than the perturbative RG-scale $\mu=40$\,GeV used in \cite{Gao:2021wun} for precision computations in the vacuum, but suffices for the present accuracy goals. We use
\begin{align}\nonumber
\alpha_s=&\, 0.235\,,\\[1ex]
 m_{l} = &\,3.0 \,\mathrm{MeV}\,,\qquad m_s = 27 \, m_{l} = 81\,\mathrm{MeV}\,,
\end{align}
at $\mu=15$\,GeV, which is compatible with the coupling parameters in \cite{Gao:2021wun} within the same RG scheme, the MOM${}^2$-scheme.

As a benchmark result we show the light quark mass function~$M_{l}(p^2) = B_{l}(p^2)/A_{l}(p^2)$ in \Cref{fig:unquench} in comparison to the quantitative fRG-DSE results in~\cite{Gao:2021wun} and the lattice results from~\cite{Oliveira:2016muq}. From this quark propagator we compute the reduced quark condensate
\begin{align}
	\label{eq:cond}
\Delta_{l,s} = \langle\bar{q}q\rangle_{l} - \frac{m_{l}}{m_{s}} \langle\bar{q}q\rangle_{s}\,.
\end{align}
For the comparison with the lattice and functional results for the reduced condensate we have to map our present results to the respective RG-scales. This has been described in detail in \cite{Gao:2021wun} where the precision results for the quark condensates have been compared to the lattice results at the lattice RG-scale $\mu_{\textrm{lat}} = 2$ GeV. Hence we simply map the present result to the lattice RG-scale and compare it with the lattice and functional results. We are led to
\begin{align}
\Delta_{l,s}(\mu_{\textrm{lat}}) = -(277.6\,\textrm{MeV})^3\,,
\label{eq:Deltalsmulat}
\end{align}
the light chiral condensate has been computed instead of the reduced condensate. For  $\mu_\textrm{lat}$ we find
\begin{align}
	\Delta_{l}(\mu_{\textrm{lat}}) = -(274.5\,\textrm{MeV})^3\,,
	\label{eq:Deltalmulat}
\end{align}
in comparison to the the functional precision result $\Delta_{l}(\mu_{\textrm{lat}}) = (272.0\,\textrm{MeV})^3$ in~\cite{Gao:2021wun}. Another and even more direct benchmark is provided with the light quark condensate in the chiral limit: it relates to the quark mass function~\cite{Gao:2021wun,Chen:2021ikl,Williams:2006vva}, and we obtain
\begin{align}
	\Delta_{l,\chi}(\mu_{\textrm{lat}}) = -(273.9(8)\,\textrm{MeV})^3\,,
	\label{eq:Deltalchimulat}
\end{align}
in comparison with the functional precision result in the vacuum  $(269.3(7) \textrm{MeV})^3$~\cite{Gao:2021wun}, and the lattice result $\Delta_{l} = (272(5)\,\textrm{MeV})^3$ (FLAG~\cite{FlavourLatticeAveragingGroup:2019iem}). Moreover, using the Pagels-Stokar formula~\cite{Gao:2020qsj} (PS) we obtain an estimate for the pion decay constant of $f_{\pi} = 94.7$ MeV. Given the expected $~10\%$ accuracy of the PS result from the full results this agrees well with $f_\pi\approx 93$\,MeV.  Moreover, the Gell-Mann--Oakes--Renner relation yields a pion mass of $m_{\pi} = 140.4$ MeV.

In summary, despite its relative simplicity the quark propagator and the derived observables in the vacuum, obtained from the present approximation show an already impressive agreement with the precision functional results and those from lattice simulations. Finally, we note that the truncation scheme is free from any phenomenological parameter, which will also be the case when applied at finite temperature and chemical potential in the following Sections.

%%%%%%%%%%%%%%%%%%%%%%%%%%%%%%%%%%%%%%
\section{ QCD phase structure}
\label{sec:PhaseStructure}

In this section we discuss the remaining details of the miniDSE scheme at finite temperature and density. This concerns in particular the thermal split and the treatment of the gluon sector. Then the phase structure of QCD is computed and confronted with that obtained with lattice simulations and functional approaches at vanishing density and functional approaches at finite density. The latter results offer a quantitative benchmark up to densities $\mu_B/T\lesssim 4$.

%%%%%%%%%%%%%%%%%%%%%%%%%%%%%%%%%%%%%%
\subsection{miniDSE scheme at finite $T$ and $\mu_B$}
\label{sec:miniDSEmuT}

\begin{figure}[b]
	\centering
	\includegraphics[width=1.0\columnwidth]{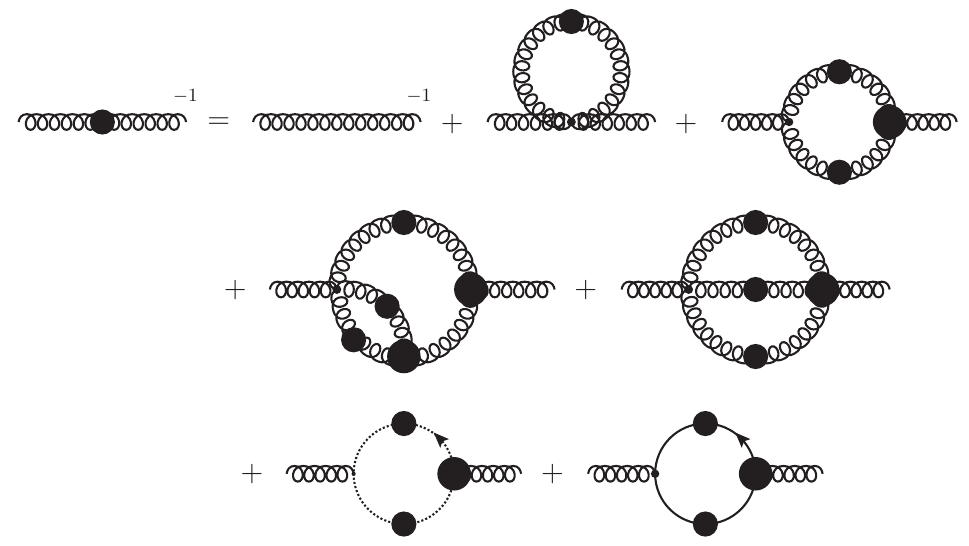}
	\caption{A diagrammatic depiction of the gluon DSE.}
	\label{fig:GluonDSE}
\end{figure}

The full quark and gluon propagators $S(p)$ and $D_{\mu\nu}(p)$ at finite temperature and density are paramterised as follows,
\begin{align}\nonumber
S^{-1}(\tilde p) = i\gamma_4\tilde{\omega}_n \, C(\tilde p) + i\boldsymbol{\gamma}\cdot\boldsymbol{p} \, A(\tilde p) + B(\tilde p)\,, \\[1ex]
p^2 \, D_{\mu\nu}(p) = \Pi_{\mu\nu}^{\textrm{E}}(p) \, Z_\textrm{E}(p) + \Pi_{\mu\nu}^{\textrm{M}}(p)  \, Z_\textrm{M}(p)\,,
\label{eq:PropsmuT}
\end{align}
with
\begin{align}
	\tilde{\omega}_n = \omega_n + i\mu_q\,,\quad  \tilde p=p+i\mu_q\,,\quad p=(\boldsymbol{p},\omega_n)\,,
	\label{eq:omega+mu}
\end{align}
and the quark Matsubara frequencies $\omega_n = (2n+1)\,\pi T$ and the gluon Matsubara frequencies $\omega_n = 2n\,\pi T$.  \Cref{eq:PropsmuT} also depends on the electric and magnetic gluon projection operators $P_{\mu\nu}^{\textrm{E,M}}$,
\begin{align}\nonumber
  \Pi_{\mu\nu}^{\textrm{M}}(p) = &\,(1-\delta_{\mu 4})(1-\delta_{\nu 4}) \left( \delta_{\mu\nu} - \frac{p_{\mu} p_{\nu}}{\boldsymbol{p}^2} \right)\,, \\[1ex]
  \Pi_{\mu\nu}^{\textrm{E}}(p) = &\,\delta_{\mu\nu} - \frac{p_{\mu} p_{\nu}}{p^2} - \Pi_{\mu\nu}^{\textrm{M}}\,.
\end{align}
The quark DSE at finite $(T,\mu_B)$ is of the form \labelcref{eq:QuarkDSE} with a spatial momentum integral and a thermal sum over Matsubara frequencies,
\begin{align}
	\sumint\limits_q  = T\sum_{n\in \mathbbm{Z}} \int \frac{d^3 q}{(2 \pi)^3}\,.
\end{align}
The DSE of the gluon propagator at finite $T$ and quark chemical potentials $(\mu_u,\mu_d,\mu_s)$ is computed along the lines suggested in \cite{Gao:2020qsj}. A diagrammatic depiction of the gluon DSE is provided in \Cref{fig:GluonDSE}. %\cJMP{Diagrammatic depiction}.
We first use the difference DSE for the gluon propagator as in \labelcref{eq:DeltaS,eq:DeltaGammamu} in an expansion about the gluon propagator in the vacuum,
\begin{align}
D_{\boldsymbol{v}}^{-1}(k) = D_{\boldsymbol{v}'}^{-1}(k) + \Delta \Pi_{A;\boldsymbol{v},\boldsymbol{v}'}(k)  \,,
\label{eq:DeltaGluon}
\end{align}
with
\begin{align}
	\boldsymbol{v}=(N_f,m_f,T,\mu_B)\,,\qquad  	\boldsymbol{v}'=(N_f,m_f,0,0)\,.
	\label{eq:Expansionvvprime}
\end{align}
In \labelcref{eq:DeltaGluon}, $\Pi_{A,\mu\nu}$ is the vacuum polarisation of the gluon that comprises all quantum, thermal and density fluctuations in terms of the diagrams in the DSE. In a further step we split the diagrams in the thermal and density difference DSE into the gluonic part $ \Delta \Pi_{A}^{\textrm{gl}}(k)$ whose classical three- or four-gluon vertex comes from the Yang-Mills sector, and the quark part $\Delta \Pi_{A}^{\textrm{qu}}(k)$ that is proportional to the classical quark-gluon vertex. The latter part is one-loop exact while the former one also contains two-loop diagrams.
\begin{align}
D_{\boldsymbol{v}}^{-1}(k) = D_{\boldsymbol{v}'}^{-1}(k) + \Delta \Pi_{A;\boldsymbol{v},\boldsymbol{v}'}^{\textrm{gl}}(k) + \Delta \Pi_{A;\boldsymbol{v},\boldsymbol{v}'}^{\textrm{qu}}(k)\,,
\label{eq:glqusplit}
\end{align}
The quark loop contribution $\Delta \Pi_{A}^{\textrm{qu}}$ in \labelcref{eq:glqusplit} reads
\begin{align}
\Delta \Pi_{A}^{\textrm{qu}}(k) = \sum_{f} \left[ \Pi_{A;\boldsymbol{v}}^{f}(k) - \Pi_{A;\boldsymbol{v}'}^{f}(k)\, \right],
\label{eq:DeltaPiAqu}
\end{align}
with
\begin{align}
\Pi_{A}^{f}(k) =- \frac{1}{2} Z_{1F}^{f}\, g^2 \sumint\limits_q \, \textrm{tr} \left[ \gamma_\mu S^{f}(\tilde p)\Gamma_{\nu}^{f}(k;\tilde p,\tilde q)S^{f}(\tilde q) \right]\,,
\label{eq:PiAqu}
\end{align}
for each flavour. The trace in \labelcref{eq:PiAqu} sums over Dirac indices and gauge group indices in the fundamental representation. The contribution is flavour diagonal as already assumed in the quark gap equation.

The pure gauge theory part can be evaluated analogously. While the difference does not require renormalisation, the numerical implementation of this property requires some care and for this purpose a numerically stable scheme has been set up and successfully used in~\cite{Gao:2020qsj, Gao:2020fbl}. In the present work we resort to a further simplifying approximation
suggested in ~\cite{Fischer:2014ata,Eichmann:2015kfa} and expand the gauge loop contribution in \labelcref{eq:glqusplit} about the lattice data of the Yang-Mills gluon propagator. We obtain
\begin{align}
  \Delta \Pi_A^{\textrm{gl}}(k) =&\,  \left[ D_T^{\mathrm{YM}}(k) \right]^{-1} - \left[ D_{T=0}^{\mathrm{YM}}(k) \right]^{-1} \,,
  \label{eq:glDSE2}
\end{align}
where we have used that YM theory is only sensitive to the temperature and not the rest of the parameters in $\boldsymbol{v}$ and $\boldsymbol{v}'$. The systematic error of this approximation for physical quark masses has been evaluated in detail in \cite{Fu:2019hdw} and does not add significantly to the total systematic error for the $T,\mu_B$ regime considered here. In a forthcoming work this approximation is also resolved with the numerically stable scheme from~\cite{Gao:2020qsj, Gao:2020fbl}.

Finally, we have to consider thermal and density splits in the vertices and especially in the quark-gluon vertex. The miniDSE approximation of the latter with two tensor structures has been  introduced in \Cref{sec:miniDSE} in the vacuum, see \labelcref{eq:Minimalqgvertex}. At finite $T,\mu_B$ we  have to take into account the thermal or density split of tensor structures as the heat bath or medium singles out a rest frame. To begin with, the classical tensor structure in 	\labelcref {eq:Minimalqgvertex} is split as
\begin{equation}
 \gamma_\mu \Sigma_A(q,p) \rightarrow \gamma_{\mu} \left[ \delta_{\mu 4}\, \Sigma_{C}(\tilde{q},\tilde{p}) + (1-\delta_{\mu 4}) \, \Sigma_{A}(\tilde{q},\tilde{p}) \right]\,,
\label{eq:STI_FT}
\end{equation}
where $\tilde q,\tilde p$ contain complex frequencies \labelcref{eq:omega+mu}. The vertex part with the second tensor structure ${\cal T}^{(4)}$ in 	\labelcref{eq:Minimalqgvertex} is split as follows,
\begin{align}\nonumber
&{\cal T}^{(4)}(q,p) \,\lambda^{(4)}(q,p)  \\[1ex]
\to &	{\cal T}^{(4)}(p,q) \,\Bigl[  \Pi^{\textrm{E}}(k)  \lambda^{(4)}_\textrm{E}(q,p)  +  \Pi^{\textrm{M}}(k)   \lambda^{(4)}_\textrm{M}(q,p) \Bigr] \,,
\end{align}
with the miniDSE approximation for the electric and magnetic dressing functions
\begin{align}
\lambda_4^\textrm{E,M}(k;\tilde{q},\tilde{p}) = g_s \,Z_\textrm{E,M}^{-1/2}(k^2) \, \Delta_B (\tilde{q},\tilde{p}).
	\label{eq:L4_FT}
\end{align}
This concludes the discussion of the simplified version of the miniDSE scheme used in the present work: we have reduced the task of solving the gap equations and vertex DSEs to that of solving the gap equations, where each approximation step has been benchmarked and controlled by functional results obtained within more sophisticated approximations as well as lattice results. We proceed by solving this set of difference DSEs for the quark and gluon dressings with the coupled quark and gluon DSEs \labelcref{eq:QuarkDSE,eq:DeltaGluon}.

%%%%%%%%%%%%%%%%%%%%%%%%%%%%%%%%%%%%%
\subsection{Chiral phase structure}
\label{sec:ChiralPhaseStructure}

We now present results for the chiral phase structure of QCD obtained in the isospin-symmetric  approximation and with a vanishing strange quark chemical potential, $(\mu_u,\mu_d,\mu_s) = (1/3 \,\mu_B, 1/3 \,\mu_B, 0)$, which give the net-baryon number density $n_B = 2/3 \,n_{u,d}$ and the vanishing strange quark density $n_s = 0$. This matches the scenario of heavy-ion collision with a vanishing net strangeness.

%------chi_T-------
\begin{figure}[t]
  \centering
  ~\includegraphics[width=0.45\textwidth]{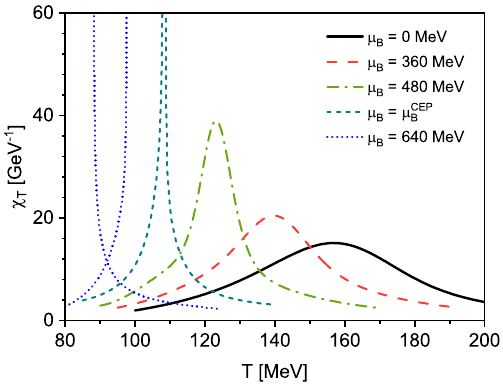}
  \caption{Temperature dependence of the susceptibility $\chi_T$ at several quark chemical potentials, including $\mu_B^{\textrm{CEP}} = 567$ MeV. For the case of $\mu_B = 640$ MeV, first order phase transition occurs and the $\chi_T$ for both Nambu and Wigner solutions are shown.}\label{fig:suscTmu}
\end{figure}
%--------------
We define the pseudo-critical temperature of the chiral phase transition $T_c(\mu_B)$ the peak temperature of the thermal susceptibility of the reduced condensate $\Delta_{l,s}$ defined in \labelcref{eq:cond},
\begin{align}
  \chi^{\ }_T(T,\mu_B) = -\partial _T \left( \frac{\Delta_{l,s}(T,\mu_B)}{\Delta_{l,s}(0,0)} \right)\,.
  \label{eq:ThermalSusceptDelta}
\end{align}
Numerical results of $\chi_T$ at several chemical potentials are shown in \Cref{fig:suscTmu}. At zero $\mu_B$, we obtain $T_c(0) = 156.5$ MeV  in agreement with results from lattice QCD~\cite{Borsanyi:2020fev,HotQCD:2018pds,Bonati:2018nut} and functional approaches~\cite{Fischer:2014ata,Fu:2019hdw,Gao:2020qsj,Gao:2020fbl,Gunkel:2021oya}.

A further benchmark result is provided with the curvature coefficients of the pseudo-critical temperature at $\mu_B=0$. Its Taylor at $\mu_B=0$ is given by
\begin{align}\label{eq:Tcmu}
  \frac{T_c(\mu_B)}{T_c(0)} = 1 - \kappa_2 \left(\frac{\mu_B}{T_c(0)} \right)^2 - \kappa_4 \left(\frac{\mu_B}{T_c(0)} \right)^4 + \cdots\,,
\end{align}
and the present simplified version of the miniDSE scheme yields
\begin{align}
\kappa_2 = 0.0169(6)\,.
\label{eq:kappa}
\end{align}
This result is slightly larger but compatible with lattice QCD~\cite{Cea:2014xva,Borsanyi:2020fev,HotQCD:2018pds} and  fRG/fRG-DSE~\cite{Fu:2019hdw,Gao:2020qsj,Gao:2020fbl} predictions with $\kappa_2 \approx 0.015$ ($0.0142(2)$ in \cite{Fu:2019hdw}, $0.0147(5)$ in \cite{Gao:2020fbl}). On the other hand, we found $\kappa_4 \approx 5\times 10^{-4}$ which is also larger but of the same magnitude as the functional results $\kappa_4 \approx 3\times 10^{-4}$ in quantitative approximations~\cite{Gao:2020fbl}.
\begin{figure}[t]
	\centering
	~\includegraphics[width=0.45\textwidth]{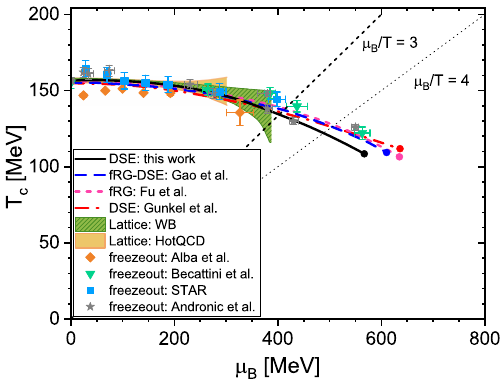}
	\caption{Phase diagram obtained here within the miniDSE scheme, compared to other functional QCD studies~\cite{Gao:2020qsj, Gunkel:2021oya, Fu:2019hdw}, lattice QCD extrapolation~\cite{Borsanyi:2020fev,HotQCD:2018pds}, and the extracted freeze-out data from different groups~\cite{Alba:2014eba,Becattini:2016xct,STAR:2017sal,Andronic:2017pug}. The present approximation to the minimal DSE scheme is reliable up to $\mu_B/T\lesssim 3$, which is marked by the black dashed line. We also display the reliability of the full quantitative computations~\cite{Gao:2020fbl,Fu:2019hdw}, the dotted line with $\mu_B/T\lesssim 4$.  }\label{fig:PD}
\end{figure}
%

%-----EoS at finite $T$ and $\mu_B$-----
\begin{figure*}[t]
	\centering
	\includegraphics[width=0.32\textwidth]{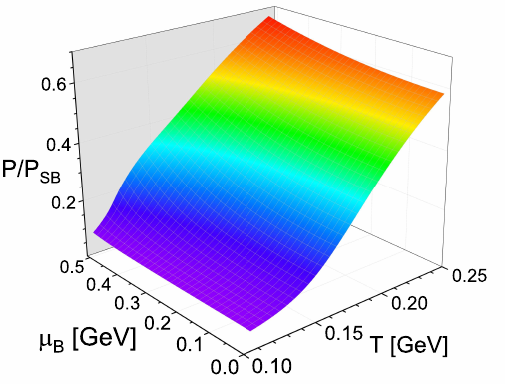}
	\includegraphics[width=0.32\textwidth]{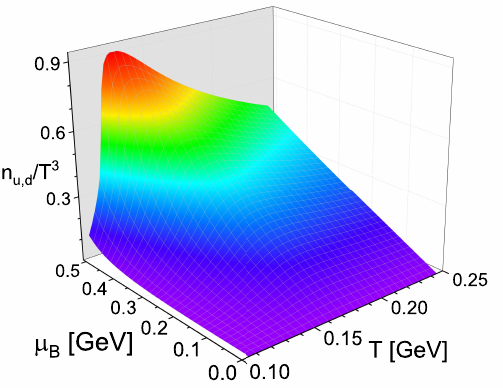}
	\caption{Light quark number density $n_{u,d}$ and QCD pressure $P$, normalised by the Boltzmann limit~\labelcref{eq:SBlimits}, at finite temperature $T$ and baryon chemical potential $\mu_B$. }
	\label{fig:eos}
\end{figure*}
\begin{figure*}[t]
	\includegraphics[width=0.32\textwidth]{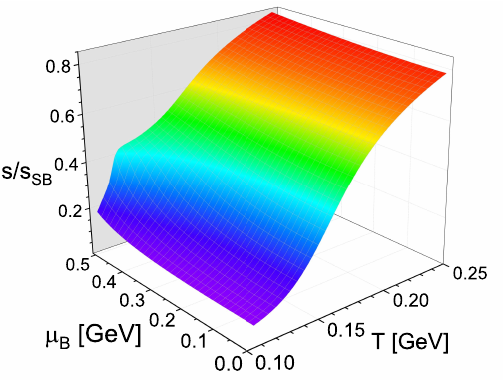}
	\includegraphics[width=0.32\textwidth]{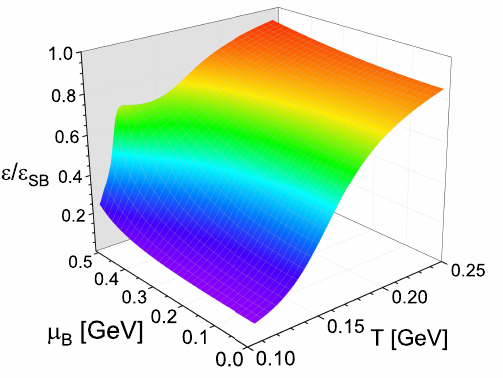}
	\includegraphics[width=0.32\textwidth]{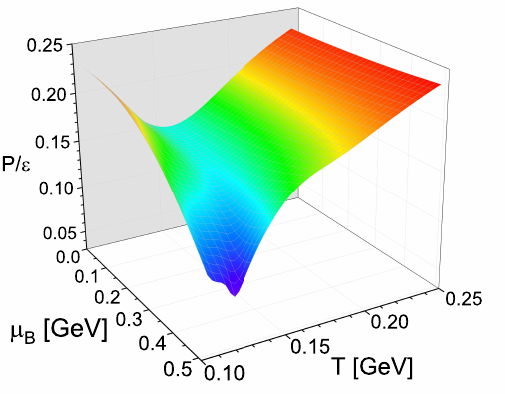}
	\caption{Calculated entropy density $s$, energy density $\epsilon$, which are scaled by their Boltzmann limits~\labelcref{eq:SBlimits}, and the ratio of pressure to energy density $P/\epsilon$.}
	\label{fig:eos_1}
\end{figure*}
%---------------------
These slight deviation grow larger at finite chemical potential. In \Cref{fig:PD} we depict the  obtained phase transition line in \Cref{fig:PD} in comparison to other functional and lattice studies.  Our result agrees well with the previous functional QCD results within more sophisticated truncations till  $\mu_B \approx 400$ MeV or $\mu_B/T \approx 3$. For $\mu_B/T \gtrsim 3$ the deviations become sizable, which also manifests itself in the location of the critical end point (CEP) with
\begin{align}
(T^{\textrm{CEP}},\mu_B^{\textrm{CEP}}) = (108.5,567)\,\textrm{MeV}\,.
\label{eq:CEPminiDSE}
\end{align}
This location has to be contrasted with the quantitative estimate
\begin{align}
(T^{\textrm{CEP}},\mu_B^{\textrm{CEP}}) \approx  (100-110\,,\,600-650)\,\textrm{MeV}\,.
\label{eq:CEPestimate}
\end{align}
from the results in \cite{Fu:2019hdw, Gao:2020fbl, Gao:2020qsj}. Note that \labelcref{eq:CEPestimate} singles out a line and not an area. In short, \labelcref{eq:CEPminiDSE} shows a $\sim 10\%$ deviation with respect to the estimate \labelcref{eq:CEPestimate} and this deviation provides a systematic error estimate for the simplified miniDSE scheme used in the present work. In summary, this analysis entails that the simplified miniDSE scheme, provides semi-quantitative results for a large range of chemical potentials. Hence, we can use it for the computation of thermodynamic quantities which are directly related to the measurements.

We close this Section with a brief discussion of the twofold origin of the deviations, that are responsible for a successive loss of fully quantitative reliability of the present results for $\mu_B/T\gtrsim 3$. To begin with, we already know from the comparison of the phase structure computation in \cite{Gao:2020qsj}, that the use of full vacuum dressings for the quark-gluon vertex corrects the curvature coefficient $\kappa$. Moreover, the deviation at larger chemical potential is also caused by the use of $\Delta_B$, \labelcref{eq:DeltaB}, in the dressing $\lambda^{(4)}$, \labelcref{eq:lambda_4}: in comparison to the dressing computed in \cite{Gao:2020fbl}, $\Delta_B$ carries a singular momentum dependence. This can be compensated for with the introduction of higher order corrections from the scattering kernel together with the imaginary part of the propagator induced by the chemical potential. An upgrade of the present simplified miniDSE scheme based on two-point dressings is work in progress and we hope to report on the respective results soon.

Another interesting aspect is the negligible contribution of the thermal chemical potential splits. For example, we find that the difference of  chiral crossover temperature for the $O(4)$-symmetric vertex without split and the vertex with thermal split is less than 1 MeV, and the curvature is barely changed. This results is also corroborated within a DSE computation with full vertices, \cite{LGPS:2023} as well as many fRG tests, see e.g.~\cite{Dupuis:2020fhh}. In conclusion, the split affects mainly the quark and gluon propagators, and the $O(4)$-symmetric approximation for the quark-gluon vertex gives agreeing results for $\mu_B/T\gtrsim 3$ as discussed above. Note however, that the explicit results here are obtained within the thermal split.

%%%%%%%%%%%%%%%%%%%%%%%%%%%%%%%
\section{Equation of state of QCD}
\label{sec:EoS}

The miniDSE scheme allows for a numerically cheap complete scan of the  EoS and other observables in the phase diagram of QCD. The quark number densities $n_q^f$ are directly obtained from the quark propagators,
\begin{align}
 n_{q}^{f}(T,\mu_B)  \simeq   -N_{c} Z_{2}^{f}\,T \, \sum_{n} \int \! \frac{\mathrm{d}^3 p
 	}{(2\pi)^3} \textrm{tr}_{D}\left[\gamma_{4}^{} S^{f}(p)\right]\,,
\label{eq:nq}
\end{align}
where we use $\mu_B= 3 \mu_l$ with $\mu_u=\mu_d=\mu_l$ and $\mu_s=0$. In the present work we simply use the momentum-dependent propagators in the $T-\mu_B$ plane on the right hand side of \labelcref{eq:nq} and leave a more detailed analysis to future work:\\[-1ex]

Firstly, it is well-known that \labelcref{eq:nq} has to be evaluated in the non-vanishing background $\langle A_0\rangle $ that solves the equations of motion, see~\cite{Braun:2007bx, Fister:2013bh}. This is tantamount to implementing the non-trivial expectation value of the Polyakov loop away from unity. Only with such a background the change from quark-gluon degrees of freedom to hadronic ones is described accurately. This is well illustrated with the kurtosis whose asymptotic temperature values is $1/9$ in the quark-gluon phase for large temperatures and unity in the hadronic phase for vanishing temperature, capturing the change of the degrees of freedom from asymptotically free quarks to weakly interacting baryons. Without the $A_0$ background the degrees of freedom in the low temperature phase resemble the quarks and the kurtosis is far smaller than unity,  for a detailed discussion see \cite{Fu:2015naa}. In short, with $\langle A_0\rangle =0$ the qualitative behaviour around the crossover line with its change of the dynamical degrees of freedom is captured, for the quantitative or even semi-quantitative behaviour the $A_0$-background is required. Respective results and formal developments in functional approaches can be found in \cite{Fu:2015amv, Fu:2016tey, Fu:2019hdw, Fu:2021oaw, Fu:2023lcm, Fu:2023lcm}. \\[-1ex]

Secondly, the density \labelcref{eq:nq} requires renormalisation and is subject to a non-trivial normalisation, reflecting its UV degree of divergence. This intricacy worsens at large temperatures but can be resolved by representing the density in terms of a (multiple) chemical potential integration of density fluctuations with a lower or absent UV degree of divergence, e.g.~the kurtosis. Indeed, the thermodynamic relation between pressure and quark number density discussed below is precisely of this type as the quark number density has a lower UV degree of divergence. \\[-1ex]

Both issues will be addressed in a forthcoming work and we proceed with the present qualitative approximation. The EoS follows from $n_q^f$ in the $(T,\mu_B)$ plane with the thermodynamic relation between the pressure and quark number densities,
\begin{align}
P(T,\mu_B) = P(T,0) + \frac13 \int_{0}^{\mu_B}  \mathrm{d}\mu\, n_{l}(T,\mu) \,,
\label{eq:prs}
\end{align}
where $n_l=n_q^u+n_q^d$. The standard thermodynamic relation \labelcref{eq:prs} is of the same structural form as our difference DSE: the integral in \labelcref{eq:prs} is simply $\Delta P(T,\mu_B) = P(T,\mu) -P(T,0)$ and follows from the quark propagators. In turn, the pressure at vanishing chemical potential can be determined from the QCD trace anomaly $I(T)$
\begin{align}
	 I(T) = (\epsilon-3P)/T^4\,,
	\end{align}
with
\begin{align}
  P(T,\boldsymbol{0})/T^4 = \int_0^T \, \mathrm{d}T' (I(T')/T')\,.
\label{eq:TA}
\end{align}
For $I(T)$ we use 2+1 flavor QCD lattice data~\cite{Borsanyi:2010cj}. Moreover, the integral over the quark number density expresses the density part of the pressure in terms of a less-divergent operator which stabilises the numerical computation and lowers the systematic error.

In summary, with the lattice input for the trace anomaly at $\mu_B=0$ and the relations \labelcref{eq:nq,eq:prs}, we can compute the QCD pressure $P(T,\mu_B)$, the energy density $\epsilon$ and the  entropy density $s$ and in the $T$-$\mu_B$ plane,
\begin{align}\nonumber
  \epsilon(T,\mu_B) = &\,T \,s(T,\mu_B) + \mu_B \,n_B(T,\mu_B) - P(T,\mu_B)\,,  \\[1ex]
  s(T,\mu_B) = &\,\partial P(T,\mu_B) / \partial T\,.
\label{eq:thermal}
\end{align}
The respective numerical results for the pressure $P/P_\textrm{SB}$ and the light quark number density $n_{u,d}/T^3$ are shown in \Cref{fig:eos} and provide us with the EoS. Further thermodynamic observables, namely the entropy density $s/s_\textrm{SB}$, the energy density $\epsilon/\epsilon_\textrm{SB}$ and pressure to energy density ratio $P/\epsilon$ are shown in \Cref{fig:eos_1}. We have normalised the pressure and energy density with the free Stefan-Boltzmann counter parts in three flavour QCD at zero chemical potential,
\begin{align}
  P_{\mathrm{SB}} = \frac{19}{36} \pi^2 T^4\,, \quad
  s_{\mathrm{SB}} = \frac{19}{9} \pi^2 T^3\,, \quad
  \epsilon_{\mathrm{SB}} = \frac{19}{12} \pi^2 T^4\,.
\label{eq:SBlimits}
\end{align}
In the vicinity of the CEP, the entropy $s$ and the energy density $\epsilon$ experience rapid changes close to the chiral crossover line $T_c(\mu_B)$. This rapid change indicates the increasingly rapid change of the degrees of freedom from hadrons to quarks in the vicinity of crossover. Moreover, the successively sharper and deeper minimum  of $P/\epsilon$ is related to the peak of the trace anomaly in \labelcref{eq:TA} as well as the minimum  of the speed of sound, and leaves a strong imprint on the EoS. The latter allows us to estimate the location of the CEP even relatively far away from it.

%---isentropic----
\begin{figure}[t]
  \centering
  \includegraphics[width=0.45\textwidth]{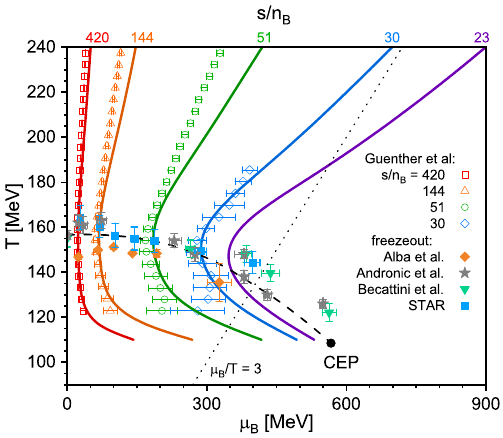}
  \caption{Isentropic trajectories for several values of $s/n_B$ together with QCD phase diagram; the black-dashed curve stands for the chiral crossover phase transition line. The trajectories  are consistent with the lattice QCD calculation as shown with the open points~\cite{Guenther:2017hnx}.  The filled points mark the freeze-out points from Refs.~\cite{Alba:2014eba,Becattini:2016xct,STAR:2017sal,Andronic:2017pug}.
  }\label{fig:isen}
\end{figure}
%------------

We have also investigated the isentropic trajectories, i.e. the trajectories satisfying $s/n_B=\textrm{const.}$ in the $(T,\mu_B)$ plane, which are related to the cooling of the hot QGP matter produced in heavy-ion collision experiments.
The isentropic trajectories calculated from our EoS at these $s/n_B$ values are shown in \Cref{fig:isen}, together with the chiral phase transition line and the CEP. We also compare the obtained phase diagram and the trajectories to the freeze out data, which are marked with the same labels as in \Cref{fig:PD}. In the vicinity of the phase transition line,
our calculated trajectories are in good agreement with those obtained from the state-of-the-art equation of state NEoS in \cite{Guenther:2017hnx,Bollweg:2022fqq}. Especially,  our trajectories for $s/n_B = 420$, 144, 51 and 30 which values are chosen in the previous studies for the corresponding collision energies in heavy ion collision experiments,    also precisely  meet  with the freeze-out points at $\sqrt{s_{\textrm{NN}}} = 200$, 62.4, 19.6 and 11.5 GeV, respectively.

At high temperatures, our results deviate from the trajectories from lattice QCD simulation and we can trace this back to the normalisation intricacy of the quark number density discussed below \labelcref{eq:nq}. In turn, below the crossover line the  background $\langle A_0 \rangle$~\cite{Braun:2007bx, Fister:2013bh, Fischer:2013eca, Fischer:2014ata, Fu:2019hdw} has not been incorporated in the present computations of the density or other thermodynamic quantities and has a significant impact. A full quantitative computation is beyond the scope of the present paper and will be presented elsewhere.

In addition to the $s/n_B$-values obtained from the extrapolation of lattice data at vanishing density, we also have investigated a smaller value with $s/n_B = 23$ with the present EoS. By comparing the result with the STAR freezeout points~\cite{STAR:2017sal}, we estimate that $s/n_B = 23$ corresponds to $\sqrt{s_{\textrm{NN}}} \gtrsim 7.7 \, \textrm{GeV}$. This estimate should be taken with a grain of salt as the curve is located at the border (and beyond) the quantitative reliability regime of the present simplified miniDSE scheme, and we have neither tackled the $A_0$-background nor the normalisation issue. With this caveat we note that this trajectory still does not cross the CEP, and it may require a smaller collision energy for  approaching it.

\begin{figure}[t]
  \centering
  \includegraphics[width=0.45\textwidth]{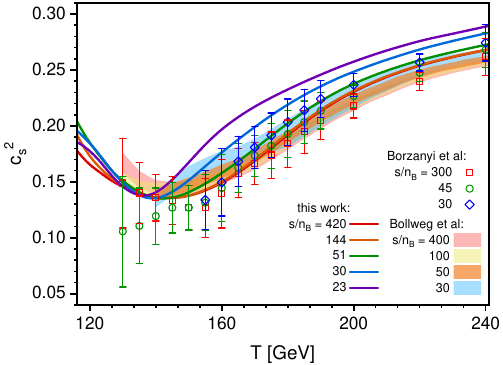}
  \caption{Speed of sound squared $c_s^2$ in  isentropic evolution as a function of temperature $T$ along the trajectories in \Cref{fig:isen}, labelled with their $s/n_B$ values. Results from the lattice calculation Ref.~\cite{Borsanyi:2012cr} and \cite{Bollweg:2022fqq} are also attached for comparison.}\label{fig:cs2}
\end{figure}
Finally, we report results for the speed of sound $c_s$ in the simplified miniDSE scheme. We have computed $c_{s}^2$ in the vicinity of phase transition line. In order to investigate the experimental scenario of adiabatic cooling, the speed of sound is evaluated along the isentropic trajectories, using the following formula~\cite{Parotto:2018pwx},
\begin{align}
  c_{s}^2
  = \frac{n_B^2 \partial_T^2 P - 2 s n_B \partial_T\partial_{\mu_B}P + s^2 \partial_{\mu_B}^2 P}{(\epsilon+P)\left[\partial_T^2 P \partial_{\mu_B}^2 P - (\partial_T\partial_{\mu_B}P)^2 \right]}.
\label{eq:cs2}
\end{align}
The temperature $T$ is chosen as the control parameter for each trajectory, and the results are shown in \Cref{fig:cs2}. The minimum of $c_s^2(T)$ agrees with the chiral phase transition point for each trajectory. The value of the speed of sound at the minimum does not change too much in the current energy range, as $c_s^2 \sim 0.13$, but the minimum becomes shaper as $s/n_B$ decreases.

The speed of sound is computed from the second and fourth order $T,\mu_B$-derivatives of QCD pressure, see \labelcref{eq:cs2}, including for example the mixed $\mu_B,T$ derivative, the thermal susceptibility of the baryon number as well as its derivative. Its minimum may be regarded as a criterion for the crossover temperature of the confinement-deconfinement phase transition. This crossover can also be measured more directly in terms of fluctuations of baryonic charges, see \cite{Fu:2023lcm} for recent functional results. We observe that the crossover temperature is a bit lower as the chiral crossover temperature defined by the peak of the thermal susceptibility of the chiral condensate, \labelcref{eq:ThermalSusceptDelta}, even though this difference does not exceed the respective error bars and the widths of these transitions. With increasing $\mu_B$ the transition regime gets sharper as the region around the minimum of $c_s^2$ is getting steeper. Hence, both the chiral and confinement-deconfinement phase transitions get steeper towards the critical end point as expected.

\begin{figure}[t]
  \centering
  \includegraphics[width=0.45\textwidth]{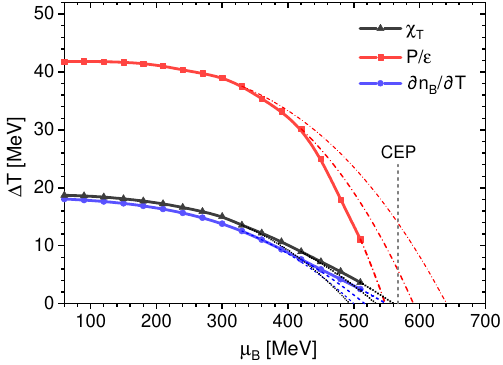}
  \caption{Estimates for the location of the critical end point from the extrapolation of the thermal width $\Delta T$, using the thermal susceptibility of the chiral condensate~$\chi_T \propto \partial_{ T} \partial_{m} P$, the thermal susceptibility of the baryon number density~$\partial n_B/\partial T = \partial_{T} \partial_{\mu_B} P$, and that of the ratio of pressure to the energy density $P/\epsilon$. The width $\Delta T$ is calculated as the width of the 90\% peak height for $\chi_T$ and $\partial n_B / \partial T$, or the 110\% value of the minimum for $P/\epsilon$. The extrapolation of $\Delta T$ is performed using the data within $\mu_B\leq330$ MeV, $420$ MeV and $510$ MeV, respectively, and the extrapolated CEP position is at $\Delta T = 0$. The actual CEP position of $\mu_B=567$ MeV is also displayed by the gray-dashed line. }\label{fig:extrapolate}
\end{figure}
Note, that we do not observe critical scaling, for a more detailed analysis see \cite{Fu:2023lcm}. However, it is precisely the smallness of the critical regime, observed by now for both the O(4)-scaling regime in the chiral limit, \cite{Braun:2020ada, Gao:2021vsf,CFG:2023} and around the critical end point~\cite{Schaefer:2011ex}, that allows for a precision estimate of the location of the latter: the extrapolation of \textit{suitable} non-universal observables towards higher chemical potentials provides a quantitative estimate of the location of the CEP, if the data are sufficiently accurate. Such an endeavour requires a theoretical search for and quantitative computation of optimal observables in the phase structure together with their extraction from high precision experimental data. A respective programme has been advocated and started in \cite{Fu:2021oaw, Fu:2023lcm} with the theoretical computation and the comparison to experimental data of fluctuations of observed charges.

In the present work we contribute to this programme by comparing the estimates of the location of the critical end point from several thermodynamic functions with the computed location in the present simplified miniDSE scheme, see \Cref{fig:extrapolate}. To that end we consider the thermal width $\Delta T$ for both thermal susceptibilities $\chi_T$ and $\partial n_B / \partial T$, which is defined as the width of the 90\% value of the peak heights of the respective susceptibility. In case of  $P/\epsilon$ the width $\Delta T$ is defined as the width of 110\% value of the minimum. These thermal widths monotonously decrease for larger chemical potential and vanish at the CEP. Hence, an extrapolation of the widths towards zero provides us with an estimate of the location of the CEP. A fully conclusive analysis will be presented elsewhere and will answer the question about the required precision and wealth of the experimental data for such a quantitative estimate in dependence of the distance to the CEP in terms of chemical potential or collision energy $\sqrt{s}$.

Here we proceed by simply elucidating this task with a limited amount of data points, see \Cref{fig:extrapolate}. We perform cubic polynomial fits for the $\Delta T$ data within several $\mu_B$ regions and then extrapolate towards larger $\mu_B$. For current $\Delta T$ data, adding higher order polynomial terms only changes the extrapolated CEP position for about 5\% and thus a cubic fit is sufficient for convergence. In the present case this originates in the sparseness of the data and not a lack of precision. We find that with successively larger $\mu_B$  included into the fit regime, the estimates for the location of the CEP  gets closer to its actual location. However, even with the present sparse data one does not have to zoom into the neighbourhood of the CEP. Moreover, the comparison shows that the chiral condensate or rather its susceptibility is better suited for such an extrapolation. In summary, it is very suggestive that a global combination of experimental precision data is best suited for such a task. This asked for the latter, which can be obtained in a combination of STAR data and in particular future high precision CBM data, based on its orders of magnitude larger luminosity.

%%%%%%%%%%%%%%%%
\section{Summary}
\label{sec:Summary}

In the present work we have computed thermodynamic quantities such as the chiral phase structure, the QCD equation of state (EoS), the isentropic trajectories and the speed of sound within first principles functional QCD. At low densities the results are benchmarked with lattice results, while at larger densities the current approach offers qualitative predictions. The EoS was obtained from integrating the quark number density from vanishing to finite chemical potential, while using lattice results for the trace anomaly at zero chemical potential as an input. Apart from the above mentioned observables we have also computed the pressure, entropy density and energy density in a wide range of temperature and chemical potential. In particular, we also discussed the implications of our results for the adiabatic speed of sound on the search for novel phases and the location of critical end point in the strong interaction matter produced in the collider experiments.

Our thermodynamic results are obtained within a minimal computational scheme for functional approaches, developed in the present work for quantitative and semi-quantitative computations, see \Cref{sec:miniDSE,sec:miniDSEmuT}. This scheme is also based on previous developments in \cite{Fu:2019hdw, Gao:2020qsj, Gao:2020fbl} both in the DSE approach as well as in the fRG approach. Here we have applied its DSE version, the miniDSE scheme, to computations of the quark propagator at finite temperature and density. Additional truncations reduced the regime of quantitative reliability to the regime $\mu_B/T \lesssim 3$, where the current results for the phase structure agree very well with that in state-of-the art quantitative truncations~\cite{Gunkel:2021oya, Fu:2019hdw, Gao:2020fbl}.
Still, also the results in the regime $\mu_B/T \gtrsim 3$ provide semi-quantitative and qualitative estimates. For example, the current estimate of the location of the critical end point only differs by approximately 10\% by that given in the quantitative studies. This leads us to the suggestion to finally determine its location within a combination of theoretical constraints and predictions for both, the phase structure as well as experimental observables, and respective experimental precision measurements.

While the current application has been tuned to minimal computational costs and further truncations have been done, aiming at the computation in terms of two-point functions alone, the fully quantitative miniDSE scheme is set-up as well. Moreover, the miniDSE scheme can also readily applied to the low temperature and finite chemical potential regime, i.e.\ cold dense quark matter and the equation of state of neutron stars. Furthermore, it provide a simple and quantitative access for the exploration of the QCD phase structure in the $(T,m_l,m_s)$ space, the Columbia plot, which is work under completion.

We hope to report soon on the respective results in the Columbia plot and for cold dense matter, and in particular on precision prediction for experimentally accessible observables in the regime $2\,\textrm{GeV} \lesssim  \sqrt{s} \lesssim 15$\,GeV. This regime includes the location of the critical end point or more generally the onset regime of new phases: Theoretical predictions accompanied with an analysis of the $\mu_B$ or $\sqrt{s}$-dependence, and a combination of STAR data and future high precision CBM data in this regime should allow us to finally pin down the location of the CEP or the onset regime of new phases as well as its physics.

%%%%%%%%%%%%%%%%%%%%%%%
\begin{acknowledgments}
	
We thank G.~Eichmann, C.~S.~Fischer, W.-j.~Fu, M.~Q.~Huber, J.~Papavassiliou, F.~Rennecke, B.-J.~Schaefer,  N.~Wink and Hui-Wen Zheng for discussions.  This work is done within the fQCD collaboration \cite{fQCD}, and we thank the members of the collaboration for discussions and collaboration on related subjects. This work is funded by the Deutsche Forschungsgemeinschaft (DFG, German Research Foundation) under Germany's Excellence Strategy EXC 2181/1 - 390900948 (the Heidelberg STRUCTURES Excellence Cluster) and the Collaborative Research Centre SFB 1225 - 273811115 (ISOQUANT). YL and YXL are supported by the National  Science Foundation of China under Grants  No. 12175007 and No. 12247107.  FG is supported by the National  Science Foundation of China under Grants  No. 12305134.

\end{acknowledgments}

\bibliography{Ref-eos}

\end{document}